# The influence of elastic strain gradient on the upper limit of flexocoupling strength, spatially-modulated phases and soft phonon dispersion in ferroics

**Anna N. Morozovska[1,2*], Christian M. Scherbakov[2], and Yulian M. Vysochanskii[3†],**

[1] *Institute of Physics, National Academy of Science of Ukraine,
46, pr. Nauky, 03028 Kyiv, Ukraine*

[2] *Taras Shevchenko Kiev National University, Physical Faculty, Chair of Theoretical Physics,
4e, pr. Akademika Hlushkova, 03022 Kiev, Ukraine*

[3] *Institute of Solid State Physics and Chemistry, Uzhgorod University,
88000 Uzhgorod, Ukraine*

**Abstract**

Within the framework of Landau-Ginzburg-Devonshire (LGD) theory we studied the role of the flexocoupling between the order parameter and elastic strain gradients in the stability of a spatially-modulated phase (SMP) in ferroics with commensurate and incommensurate long-range ordered phases under the presence of *squired elastic strain gradient*. The squired elastic strain gradient is required for the free energy stability to arbitrary strain gradients.

Obtained analytical expressions showed that the fundamental upper limit for the magnitude of the static bulk flexoelectric effect strength, established by Yudin and Tagantsev under the absence of squired elastic strain gradient and higher order gradients terms, should be substituted by the temperature-dependent condition on the flexoelectric coupling strength under the presence of the gradient terms. Moreover, we established that the SMP appears and becomes stable in commensurate ferroics if the flexocoupling constant exceeds the critical value, defined by the reduced temperature, strain and order parameter gradients constants, striction and expansion coefficients in the LGD functional.

We calculated the soft phonon dispersion in ferroics with commensurate and incommensurate long-range ordered phases allowing for the squired elastic strain gradient, as well as static and dynamic flexocoupling. Appeared that the dispersion for the optic mode is slightly sensitive to the flexocoupling, while the dispersion of acoustic mode strongly depends on the coupling strength. Obtained results demonstrate that the non-trivial differences in the dispersion of optic and acoustic modes appear under the change of flexocoupling constant. Hence the experimental determination of soft phonon dispersion can give the important information about the influence of the strain gradient and flexocoupling on the SMP in ferroics with commensurate and incommensurate long-range order. These theoretical predictions require experimental verification.

---

[*] Corresponding author 1: anna.n.morozovska@gmail.com

[†] Corresponding author 2: vysochanskii@gmail.com

**I. Introduction**

The static flexoelectric effect is an electric polarization generated in solids by a strain gradient and vice versa [1, 2]. The induced polarization component $P^{sf}$ is linearly proportional to the elastic strain gradient component $\partial u/\partial x$ and the proportionality coefficient $f$ is the corresponding component of the flexocoupling tensor. Following Kogan, $f$ value is quite small, $f \sim e/a$, where $e$ and $a$ are respectively electronic charge and lattice constant [3]. The dynamic flexoelectric effect was firstly introduced by Tagantsev as $P^{df} = -M\frac{\partial^2 U}{\partial t^2}$, where $U$ and $M$ are the components of elastic displacement and flexodynamic tensor correspondingly [4, 5], corresponds to the polarization response to accelerated motion of the medium.

Flexoelectricity occurs in all 32 crystalline point groups, because the strain gradient breaks the inversion symmetry. Owing to the universal nature, flexoelectricity permanently attracts broad scientific interest [4, 6], but its application potential in homogeneous macro-materials is fundamentally limited due to the small strength $f$. In contrast to homogeneous macroscopic systems, it is difficult to overestimate the significance of the flexoelectric phenomena in ferroics and multiferroics (e.g. antiferroelectrics, ferrielectrics, ferroelectrics, superparaelectrics and ferromagnetoelectrics) [7, 8, 9], which are either nanosized (e.g. thin films, nanoparticles, fine-grained ceramics) or possess nanoscale inhomogeneous (e.g. nanoregions, dense nanodomain structure or spatially-modulated phases) of the order parameter, such as spontaneous polarization, magnetization or antiferromagnetic order parameter [10]. The order parameter gradient interacts with elastic strain inside the nanostructured ferroic via the flexocoupling. Hence the flexo-type couplings strongly change the structural, polar and electro-transport properties of the ferroic nanoparticles [11, 12], fine-grained ceramics [13, 14], ferroelectric [15, 16, 17, 18] and ferroelastic [19, 20] domain walls and interfaces, as well as it induces reentrant phases [21] and incommensurate spatial modulation [22, 23, 24] in ferroics.

Allowing for the flexoelectricity importance for the insight to meso- and nanoscale couplings in ferroics, one has to know and separate its static and dynamic contributions. However the value of $f$ calculated theoretically [25, 26] accordingly to and Kogan's microscopic definition [3] can be of several orders of magnitude smaller than those measured experimentally [27, 28, 29]. The discrepancy motivated Yudin and Tagantsev to establish theoretically the upper limits for the magnitude of the static bulk contribution to the flexoelectric effect in ferroelectrics [30]. The obtained magnitude of the upper limit suggests that the anomalously high flexoelectric coupling measured for perovskite ceramics [27-29] can hardly be attributed to a manifestation of the static bulk effect.

Note, that Yudin and Tagantsev used Landau-Ginzburg-Devonshire (**LGD**) free energy without inclusion of the quadratic term of elastic strain gradient, $\frac{v}{2}\left(\frac{\partial u}{\partial x}\right)^2$. Rigorously speaking, the quadratic term, that is typically ignored in the free energy density of ferroics, is responsible for the stable smooth distribution of the order parameter at nonzero strain gradients, since the presence of Lifshitz type flexo-invariant, $\frac{f}{2}\left(P\frac{\partial u}{\partial x}-u\frac{\partial P}{\partial x}\right)$, essentially changes the stability conditions of the LGD free energy and elastic boundary conditions [11, 31].

Let us underline that one of the basic experimental methods collecting information about the spatial modulation of the order parameter in ferroics are dielectric measurements, neutron, Raman, and Brillouin scattering [32, 33, 34, 35, 36, 37, 38, 39]. Available experimental and theoretical results [34-39] evidently demonstrate the significant influence of the flexocoupling on the scattering spectra. Theoretical analysis of the flexoelectric coupling contribution to the soft phonon eigen vector revealed that if the stability condition limiting the upper value of the flexocoupling constant *f* becomes invalid, one can expect the appearance of the spatially-modulated phases (**SPM**) in the long-range ordered phase of a ferroelectric [40]. However the thermodynamic analysis of the SPMs stability in ferroics with commensurate phase (**CP**) and incommensurate phases (**ICP**) of the long-range order parameter under the presence of squired elastic strain gradient and higher gradients of the order parameters is absent to date.

The gap in the knowledge motivated us to study the role of the flexocoupling between the order parameter and elastic strain gradients in the stability of SMPs in ferroics with CP and ICP under the presence of squired elastic strain gradient and higher order gradients of the order parameter.

**II. Thermodynamics of the spatially-modulated phase in ferroics**

LGD free energy *F* of a ferroic acquires the simplest form for the one-component order parameter η coupled with the strain tensor component *u*, which depend on the one coordinate *x* in the one-dimensional (1D) case [35]. Full tensorial form of the free energy density that depends on three coordinates is listed in the **Appendix A** of the **Supplement**. In the simplest one-component and 1D case, considered hereinafter, the bulk part of the Helmholtz free energy *F* expansion on η, *u* and their gradients have the following form:

$$F_V = \int dx \left( \begin{array}{c} \frac{\alpha(T)}{2}\eta^2 + \frac{\beta}{4}\eta^4 + \frac{\gamma}{4}\eta^6 + \frac{g}{2}\left(\frac{\partial \eta}{\partial x}\right)^2 + \frac{w}{2}\left(\frac{\partial^2 \eta}{\partial x^2}\right)^2 + \frac{h}{2}\eta^2\left(\frac{\partial \eta}{\partial x}\right)^2 \\ -\eta E - qu\eta^2 + \frac{c}{2}u^2 + \frac{v}{2}\left(\frac{\partial u}{\partial x}\right)^2 - \frac{f}{2}\left(\eta\frac{\partial u}{\partial x} - u\frac{\partial \eta}{\partial x}\right) \end{array} \right) \quad (1)$$

According to Landau theory [41, 42], the coefficient $\alpha(T) = \alpha_T(T-T_C)$ explicitly depends on temperature *T*, $T_C$ is the Curie temperature. All other coefficients are supposed to be temperature independent. Coefficient β>0 for the ferroics with the second order phase transition and β<0 for the first order one. Nonlinear stiffness γ should be non negative (γ ≥ 0) for the functional stability. Parameters *g, w* and *v* determine the magnitude of the gradient energy. For ferroics with CP of the long-range order parameter $g>0$ and $w \geq 0$, while $g<0$ and $w>0$ for ferroics with the ICP [43]. Typically the nonlinear gradient parameter *h* is small and its influence will be neglected hereinafter. The order parameter can be conjugated with external field *E*. Depolarization/demagnetization field is regarded absent for the sake of simplicity. The striction coefficient *q* can be positive or negative. The elastic stiffness *c* and the strain gradient coefficient *v* should be always positive for the functional stability.

Coefficient *f* is the component of the static flexocoupling tensor. In fact, only the Lifshitz-type invariant $\frac{f}{2}\left(\eta_k \frac{\partial u}{\partial x} - u \frac{\partial \eta}{\partial x}\right)$ is relevant for the bulk contribution of the static flexoelectric effect. Rigorously speaking, the squired strain gradient term $\frac{v}{2}\left(\frac{\partial u}{\partial x}\right)^2$, that is typically ignored in ferroics, is responsible for the stable smooth distribution of the order parameter at nonzero strain gradients, since the presence of Lifshitz invariant essentially changes the stability conditions and elastic boundary conditions [31]. In particular, Eliseev et al [11] obtained that the term $v_{ijklmn}(\partial u_{ij}/\partial x_k)(\partial u_{lm}/\partial x_n)$ can be neglected under the condition $f_{klmn}^2 < g_{ijkl} c_{ijmn}$ in the tensorial case for a small strain gradients. However the term $\frac{v}{2}\left(\frac{\partial u}{\partial x}\right)^2$ is mandatory required for the system stability to arbitrary strain gradients.

Thermodynamic equations of state are obtained from the variation of the free energy (1) on the components of the order parameter η and strain *u*, $\delta F/\delta \eta = 0$ and $\delta F/\delta u = 0$. Their explicit form is listed in the **Appendix B** of **Supplement**. Let us find the solution of these equations after their linearization in the vicinity of spontaneous values in Fourier k-domain

$$\eta = \eta_S + \int dk \exp(ikx)\tilde{\eta}, \quad u = u_S + \int dk \exp(ikx)\tilde{u}. \quad (2)$$

Perturbation field $E = \int dk \exp(ikx)\tilde{E}$. Homogeneous spontaneous strain and order parameter values are denoted as $\eta_S$ and $u_S$ correspondingly.

In a high temperature parent phase $\alpha > 0$, $\eta_S = 0$, $u_S = 0$. In a low temperature ordered phase, where $\alpha < 0$, $\eta_S \neq 0$, $u_S \neq 0$, the homogeneous spontaneous strain and order parameter values can be determined from the equations of state at zero gradients, namely

$$u_S = \frac{q}{c}\eta_S^2, \quad \eta_S^2 = \frac{1}{2\gamma}\left(\sqrt{\beta^{*2} - 4\alpha\gamma} - \beta^*\right), \quad (3)$$

where $\beta^* = \left(\beta - 2\dfrac{q^2}{c}\right)$. Equations (3) are valid for ferroics with the first order ($\beta^* < 0, \gamma > 0$) or the second order ($\beta^* > 0, \gamma \geq 0$) phase transition to the parent phase. In particular case $\beta^* > 0, \gamma = 0$ the spontaneous value $\eta_S^2 = -\alpha/\beta^*$. Therefore the condition $c\beta > 2q^2$ should be valid for the second order phase transition realization.

Linearized solution of the equations of state has the following form:

$$\tilde{\eta} = \tilde{\chi}(k)\tilde{E}, \qquad \tilde{u} = -\dfrac{(ifk - 2q\eta_S)}{(c+vk^2)}\tilde{\chi}(k)\tilde{E}. \qquad (4)$$

The linear susceptibility (Green function) introduced in Eq.(4) has the form:

$$\tilde{\chi}(k) = \left(\alpha + 3\beta\eta_S^2 + 5\gamma\eta_S^4 - 2qu_S + gk^2 + wk^4 - \dfrac{4q^2\eta_S^2 + f^2k^2}{c+vk^2}\right)^{-1}, \qquad (5)$$

The condition of the solution (4) instability corresponds to the divergence of susceptibility (5). The instability condition, which can indicate the simultaneous appearance of the spatial modulation with a period $k$, acquires the form:

$$(\alpha_S + gk^2 + wk^4)(c+vk^2) - f^2k^2 - 4q^2\eta_S^2 = 0 \qquad (6)$$

Here the renormalized parameter $\alpha_S = \alpha + \left(3\beta - 2\dfrac{q^2}{c}\right)\eta_S^2 + 5\gamma\eta_S^4$ is introduced in the ordered phase, where $\alpha < 0$, $\eta_S \neq 0$, $u_S \neq 0$. One can show that the parameter $\alpha_S$ is always positive. The parameter $\alpha_S$ is equal to $\alpha$ in a parent phase, where $\alpha > 0$, $\eta_S = 0$, $u_S = 0$. The inhomogeneous modulated phase can appear if the renormalized gradient coefficient $g^{eff} = \left(g + \dfrac{\alpha_S v}{c} - \dfrac{f^2}{c}\right)$ becomes negative. That say, the homogeneous phase is absolutely stable under the condition $f^2 < cg + \alpha_S v$. The condition is temperature-dependent because of the temperature dependence $\alpha_S(T)$. If $v = 0$ the condition reduces to the inequality $f^2 < cg$, that is nothing more that the scalar form of tensorial relation $f_{klmn}^2 < g_{ijkl}c_{ijmn}$ suggested by Eliseev et al [11]. Later on, the condition $f_{44}^2 < g_{44}c_{44}$ (along with other similar conditions valid for perovskite symmetry) has been derived exactly by Yudin and Tagantsev [30] and presented as the upper limit for the magnitude of the static bulk contribution to the flexoelectric effect in ferroelectrics under the absence of squired strain gradient and higher order gradient of the polarization. Below we will show the necessary temperature-dependent condition $g^{eff} < 0$ per se is not sufficient for the stability of the spatial modulation.

After neglecting the smallest term $vwk^6$ Eq.(6) has the following roots:

$$k_{1,2}^{\pm} = \pm \sqrt{\frac{cg}{2(gv+cw)}\left(\frac{f^2}{cg}-1-\frac{\alpha_S v}{cg} \pm \sqrt{\left(\frac{f^2}{cg}-1-\frac{\alpha_S v}{cg}\right)^2 - 4\frac{(gv+cw)}{cg^2}\left(\alpha_S - 4\frac{q^2}{c}\eta_S^2\right)}\right)} \quad (7)$$

Introducing the dimensionless wave vector $k^*$ and parameters $a^*$, $F^*$, $\alpha_v^*$, $w^*$ and $Q^*$ in the following way:

$$k^* = \frac{ak}{\pi}, \quad a^* = \frac{a}{\pi}\sqrt{\frac{c}{2v}}, \quad F^* = \frac{f^2}{cg}, \quad \alpha_v^* = \frac{\alpha_S v}{cg}, \quad w^* = \frac{cw}{vg}, \quad Q^* = 4\frac{q^2 \eta_S^2}{c\alpha_S}, \quad (8)$$

where $a$ is a lattice constant, we can rewrite Eq.(7) in the following way:

$$k_{\pm}^* = a^*\sqrt{\frac{1}{w^*+1}\left(F^* - 1 - \alpha_v^* \pm \sqrt{(F^* - 1 - \alpha_v^*)^2 - 4\alpha_v^*(w^*+1)(1-Q^*)}\right)}. \quad (9)$$

Note, that the parameter $\alpha_v^*$ is proportional to the product of the temperature-dependent coefficient $\alpha_S$ and the strain gradient squared coefficient $v$. In particular for the ferroics with the second order phase transitions ($\beta > 0$) the square of the order parameter $\eta_S^2 = -\alpha/\beta^* \cong \alpha_T(T_C - T)/(\beta - 2q^2/c)$ and so $\alpha_S = -2\beta\alpha/(\beta - 2q^2/c) \sim \alpha_T(T_C - T)$ in the ordered phase. Hence one obtains that $\alpha_v^*$ linearly depends on temperature, $\alpha_v^* \cong 2v\alpha_T(T_C - T)/g(c\beta - 2q^2)$. After elementary transformations one obtains that the striction parameter $Q^* \cong 2q^2/(c\beta)$ is virtually temperature-independent. Hence it makes sense to study the ferroic phase diagram in dependence on the temperature-dependent parameter $\alpha_v^* \sim v\alpha_T(T_C - T)/g$ (further regarded as *reduced temperature*) and flexoelectric coupling dependent parameter $F^* \sim f^2/g$ (further regarded as *flexoconstant*). Both these parameters, which are proportional to $1/g$, are positive for ferroics with CP and negative for ferroics with ICP. The case $g = 0$ is excluded from the analyses for chosen dimensionless variables (8) assuming that we considered ferroics with composition far from the Lifshitz point.

The necessary condition of the SPM appearance, $g^{eff} < 0$, corresponds to the inequality $\text{sign}(g)(1 + \alpha_v^* - F^*) < 0$. However, as one can see from Eq.(9), the condition per se is not sufficient, because the additional condition of positive inner determinant, $(F^* - 1 - \alpha_v^*)^2 \geq 4\alpha_v^*(w^* + 1)(1 - Q^*)$, should be valid.

### III. Thermodynamic analyses of the SPM stability in ferroics with CP and ICP

Expression (9) is analyzed below for two physically different cases of CF and ICF. Since $g > 0$ for CF, they should be modeled by not negative parameters $F^* \geq 0$, $\alpha_v^* \geq 0$, $Q^* \geq 0$ and $w^* \geq 0$. For simplicity below we put $w^* = 0$ for ferroics with CP. Since $g < 0$ for ferroics with ICP, they should be modeled by not positive parameters $F^* \leq 0$, $\alpha_v^* < 0$, $Q^* \geq 0$ and $w^* < 0$.

Reference values of LDG-expansion coefficients and other material parameters can be estimated for ferroics with ICP and CP from the **Table S1** in **Appendix D** of **Supplement.** Namely $\alpha_S = (0 - 2) \times 10^3$ C$^{-2}$·mJ, β changes from $-5 \times 10^8$ J C$^{-4}$·m$^5$ for the ferroics with the first order phase transitions to $+1 \times 10^8$ J C$^{-4}$·m$^5$ for the ferroics with the second order phase transitions, γ changes from $10^9$ J C$^{-6}$·m$^9$ to $10^{11}$ J C$^{-6}$·m$^9$, $\eta_S$ changes from 0 to 0.7 C/m$^2$ in dependence on temperature, $q$ is about $10^9$ Vm/C, $c$ is about $(1 - 10) \times 10^{10}$ Pa, $g$ changes from $-6 \times 10^{-10}$ C$^{-2}$m$^3$J for ferroics with ICP like Sn$_2$P$_2$S$_6$ to $+5 \times 10^{-10}$ C$^{-2}$m$^3$J for ferroics with CP like Sn$_2$P$_2$Se$_6$, $w=0$ for ferroics with CP and is about $(1 - 3) \times 10^{-27}$ J·m$^5$/C$^2$ for ferroics with ICP; $f$ changes from $-5$ V to $+5$ V, $v$ is within the range $(10^{-7} - 10^{-6})$V s$^2$/m$^2$. Using the reference values we estimated that dimensionless parameters changes in the range $Q^* = (0.02 - 0.8)$, $w^* = 0$ for CF and $w^* = -(2 - 20)$ for ICF, $F^* = (-5 - +5)$, $v^* = (-5 - +5)$, $k^* = (0 - 0.5)$ and $a^* \sim 0.1$. Thus for a reasonable range of striction, nonlinearity and elastic stiffness coefficients the inequality is likely to be valid $0 \leq Q^* < 1$.

**III.A. Ferroics with commensurate phases.** Consequently $4\alpha_v^*(w^* + 1)(1 - Q^*) > 0$ as well as $F^* \geq 0$, $\alpha_v^* \geq 0$, $0 \leq Q^* < 1$ and $w^* \geq 0$ for ferroics with CP with $g > 0$. Using all these inequalities one can see from Eq.(8) that the homogeneous distribution of the order parameter (**HP**) in ferroics with CP is thermodynamically stable under the condition $(F^* - 1 - \alpha_v^*)^2 < 4\alpha_v^*(w^* + 1)(1 - Q^*)$, while the SMP with modulation period(s) given by Eq.(9) can appear under the conditions $(F^* - 1 - \alpha_v^*)^2 \geq 4\alpha_v^*(w^* + 1)(1 - Q^*)$ and $F^* \geq 1 + \alpha_v^*$ The necessary condition of SMP appearance is $(1 + \alpha_v^* - F^*) < 0$. Altogether these three inequalities give us the necessary and sufficient conditions of the HP and SMP stability in ferroics with CP:

$$0 \leq F^* < 1 + \alpha_v^* + 2\sqrt{\alpha_v^*(w^* + 1)(1 - Q^*)}, \quad - \quad \text{HP phase is stable,} \quad (10a)$$

$$F^* \geq 1 + \alpha_v^* + 2\sqrt{\alpha_v^*(w^* + 1)(1 - Q^*)}. \quad - \quad \text{SMP phase is stable.} \quad (10b)$$

Diagram of the SMP and PH phases existence in ferroics with CP is shown in the right-hand side of **Figure 1,** where $\alpha_v^* \geq 0$ and $F^* \geq 0$. As one can see, the *minimal positive* value of the flexoconstant $F_{cr}^*$ required for SMP appearance is $F_{cr}^* = 1$ at $\alpha_v^* = 0$. That say HP is absolutely stable at $F^* < 1$ and $\alpha_v^* = 0$. The value $F_{cr}^*$ monotonically increases with $\alpha_v^*$ increasing in accordance with the formulae $F_{cr}^*(\alpha_v^*) = 1 + \alpha_v^* + 2\sqrt{\alpha_v^*(w^* + 1)(1 - Q^*)}$.

Dependences of the wave vectors $k_+^*$ and $k_-^*$ on the dimensionless flexoconstant $F^*$ and reduced temperature $\alpha_v^*$ are shown in the right-hand side of **Figures 2a** and **2b** correspondingly. A "gap" (i.e. the absence of $k_\pm^*$) exists for all curves corresponding to different $\alpha_v^*$-values and $F^* < F_{cr}^*$ in the right-hand side of **Figure 2a**. The gap width $d$ is conditioned by the value $F_{cr}^*(\alpha_v^*)$ and it increases with $\alpha_v^*$ increasing. Equal wave vectors appear under the condition $F^* = F_{cr}^*$, $k_-^*(\alpha_v^*, F_{cr}^*) = k_+^*(\alpha_v^*, F_{cr}^*)$. At $F^* > F_{cr}^*$ the wave vector $k_-^*$ decreases and $k_+^*$ increases with $F^*$ increasing (compare dashed and solid curves in the right-hand side of **Figure 2a**). There is the *maximal positive* value $\alpha_{cr}^*(F^*)$ of SMP appearance at fixed $F^*$ value. Equal wave vectors appear under the condition $\alpha_v^* = \alpha_{cr}^*$, i.e. $k_-^*(\alpha_{cr}^*, F^*) = k_+^*(\alpha_{cr}^*, F^*)$. At $\alpha_v^* < \alpha_{cr}^*$ wave vector $k_-^*$ increases and $k_+^*$ decreases with $\alpha_v^*$ increasing (compare dashed and solid curves in the right-hand side of **Figure 2b**).

**III.B. Ferroics with incommensurate phases.** The stability conditions of the homogeneous state for ferroics with ICP are more complex, because they are different for the cases $-1 < w^* \leq 0$ and $w^* < -1$. In accordance with our estimates, the most realistic and interesting situation corresponds to the case $w^* < -1$, for which we continue our analysis. For the case $w^* < -1$ the inequality $4\alpha_v^*(w^* + 1)(1 - Q^*) > 0$ is valid. Also the inequalities $F^* \leq 0$, $\alpha_v^* < 0$ and $0 \leq Q^* < 1$ are valid for ferroics with ICP. Using all these inequalities one can see from Eq.(9) that the HP is thermodynamically stable for ferroics with ICP under the same condition $(F^* - 1 - \alpha_v^*)^2 < 4\alpha_v^*(w^* + 1)(1 - Q^*)$ as for ferroics with CP, while the SMP with modulation period(s) given by Eq.(9) can appear under the conditions $F^* \leq 0$, $F^* \leq 1 + \alpha_v^*$ and $(F^* - 1 - \alpha_v^*)^2 \geq 4\alpha_v^*(w^* + 1)(1 - Q^*)$. The necessary condition of SMP appearance is $(1 + \alpha_v^* - F^*) < 0$. Altogether these inequalities give us the necessary and sufficient conditions of the HP and SMP stability in ferroics with ICP:

$$1 + \alpha_v^* - 2\sqrt{\alpha_v^*(w^* + 1)(1 - Q^*)} < F^* < 0, \quad - \text{ HP phase is stable,} \qquad (11a)$$

$$F^* \leq 0 \text{ and } F^* \leq 1 + \alpha_v^* - 2\sqrt{\alpha_v^*(w^* + 1)(1 - Q^*)}. \quad - \text{ SMP phase is stable.} \qquad (11b)$$

The diagram of the SMP and PH phases existence in ferroics with ICP is shown in the left-hand side of **Figure 1**, where $\alpha_v^* \leq 0$ and $F^* \leq 0$. In contrast to ferroics with CP, SMP appears at $F^* = 0$ and $\alpha_v^* \geq -0.3$. Then the *maximal negative* value $F_{cr}^*(\alpha_v^*)$ appears and its absolute value increases quasi-linearly with $|\alpha_v^*|$ increasing in accordance with the formulae $F_{cr}^*(\alpha_v^*) = 1 + \alpha_v^* - 2\sqrt{\alpha_v^*(w^* + 1)(1 - Q^*)}$.

Dependences of the wave vectors $k_+^*$ and $k_-^*$ on the flexoconstant $F^*$ and reduced-temperature $\alpha_v^*$ are shown in the left-hand side of **Figures 2a** and **2b** correspondingly. A gap $d$ exists only for the two curves corresponding to the highest values of $|\alpha_v^*|$, as it follows from the dependence $F_{cr}^*(\alpha_v^*)$ (see the left-hand side of **Figures 2a**). Wave vectors are equal at $F^* = F_{cr}^*$, $k_-^*(\alpha_v^*, F_{cr}^*) = k_+^*(\alpha_v^*, F_{cr}^*)$. At $F^* < F_{cr}^*$ the wave vector $k_-^*$ decreases with $|F^*|$ increasing, while $k_+^*$ increases with $|F^*|$ increasing (compare dashed and solid curves in left-hand side of **Figure 2a**). There is the *minimal* negative value $\alpha_{cr}^*(F^*)$ of SMP appearance at fixed $F^*$ value. Wave vectors are equal at the value $\alpha_v^* = \alpha_{cr}^*$, $k_-^*(\alpha_v^*, F^*) = k_+^*(\alpha_v^*, F^*)$. At $|\alpha_v^*| < |\alpha_{cr}^*|$ wave vector $k_-^*$ increases and $k_+^*$ decreases with $|\alpha_v^*|$ increasing (compare dashed and solid curves in the right-hand side of **Figure 2b**).

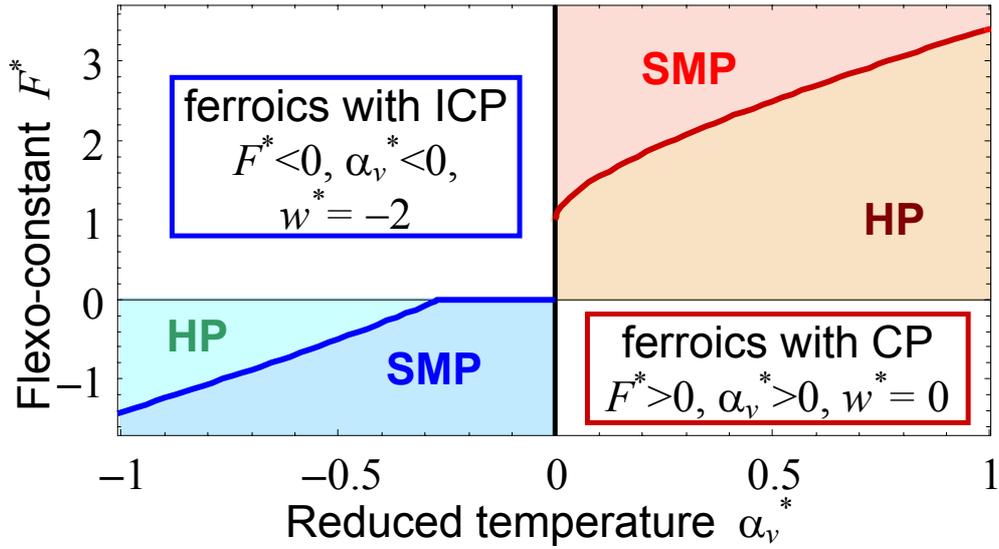

**Figure 1.** Diagram of the spatially-modulated phase (SMP) and homogeneous phase (PH) stability in ferroics with CP (right side) and in ferroics with ICP (left side) plotted in dimensionless coordinates, flexoconstant $F^*$ and reduced-temperature $\alpha_v^* \sim v\alpha_T(T_C - T)$. Parameter $Q^* = 0.5$; parameter $w^* = -2$ ferroics with ICP and $w^* = 0$ for ferroics with CP.

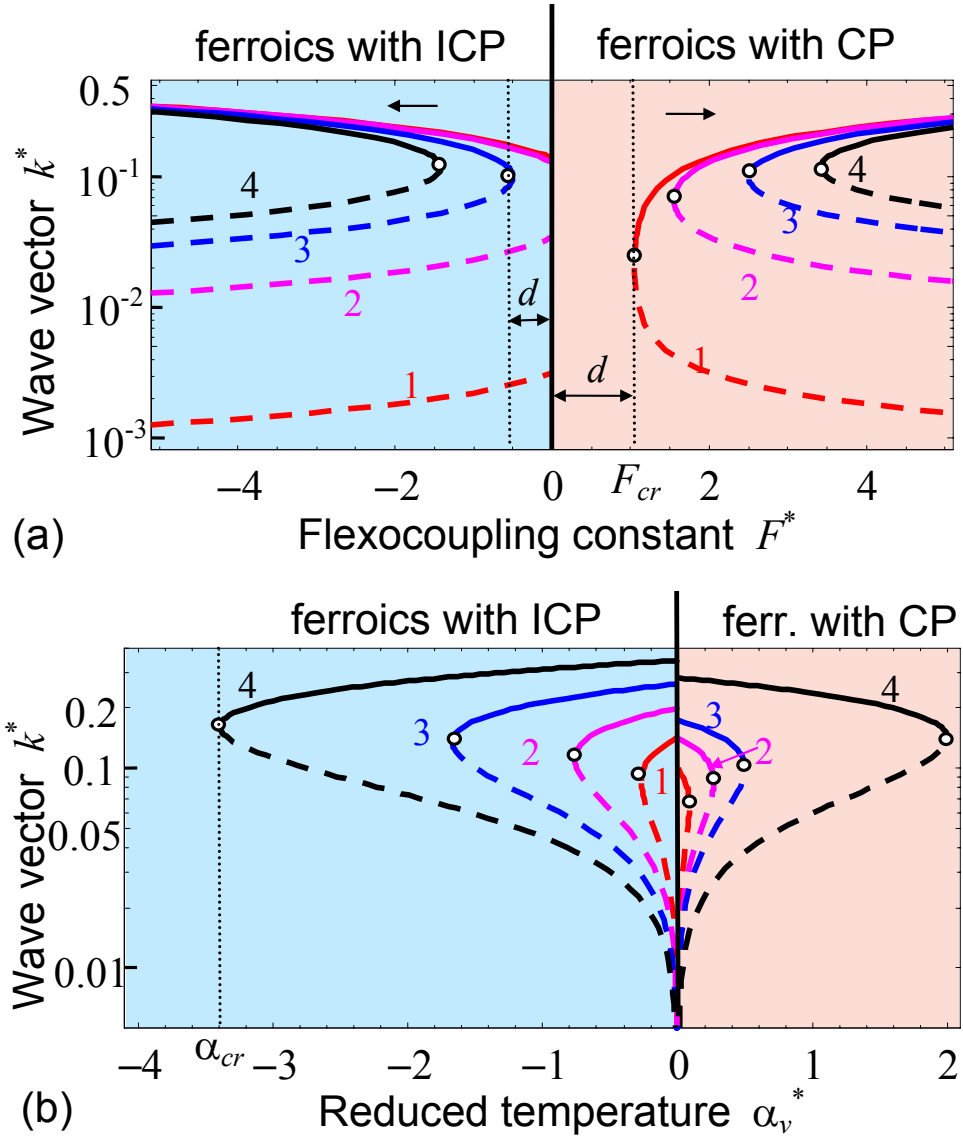

**Figure 2.** Dependences of the wave vectors $k_+^*$ (solid curves) and $k_-^*$ (dashed curves) on the flexoconstant $F^*$ (a) and reduced-temperature $\alpha_v^*$ (b). Different curves are calculated for several different values of $\alpha_v^* = 0.001$, 0.1, 0.5, 1 for ferroics with CP and $\alpha_v^* = -0.001, -0.1, -0.5, -1$ for ferroics with ICP (curves 1-4 in the plot (a)); $F^* = 1.5, 2, 2.5, 5$ for ferroics with CP and $F^* = -0.001, -1, -2.5, -5$ for ferroics with ICP (curves 1-4 in the plot (b)) listed in the legends at the plots. Parameters $Q^* = 0.5$ and $a^* = 0.1$; parameter $w^* = -2$ for ferroics with ICP and $w^* = 0$ for ferroics with CP.

To resume, performed analytical analyses showed that the fundamental upper limit for the magnitude of the static bulk contribution to the flexoelectric effect existing under the absence of the squared strain gradient and higher order gradient terms [30] should be substituted by the temperature-dependent conditions (10)-(11) corresponding to the HP stability or SPM phase appearance.

## IV. Soft phonon dispersion in ferroics with CP and ICP

Soft phonon dispersion can be calculated from the time-dependent dynamic equations of state for the order parameter $\eta$ and elastic displacement component $U$, $\delta L/\delta\eta = 0$ and $\delta L/\delta U = 0$ [40], which explicit form in listed in the **Appendix C** of **Supplement**. Lagrange function $L = \int_t dt(F - E)$ consists of the free energy $F$ given by Eq.(1) and kinetic energy $E$ that is given by expression

$$E = \int dx \left( \frac{\mu}{2}\left(\frac{\partial\eta}{\partial t}\right)^2 + M\frac{\partial\eta}{\partial t}\frac{\partial U}{\partial t} + \frac{\rho}{2}\left(\frac{\partial U}{\partial t}\right)^2 \right), \quad (12)$$

which includes the dynamic flexoelectric coupling [4, 5] with the magnitude $M$, $\rho$ is the density of a ferroic, elastic displacement component $U$ is related with the strain $u$ as $u = \partial U/\partial x$.

The solution of dynamic equations was found after their linearization in the vicinity of spontaneous values $\eta = \eta_S + \int dk \exp(ikx + i\omega t)\tilde{\eta}$ and $U = u_S x + \int dk \exp(ikx + i\omega t)\tilde{U}$ in k-$\omega$ Fourier domain. Perturbation field $E = \int dk \exp(ikx + i\omega t)\tilde{E}$ (see **Appendix C** of **Supplement**). The linear susceptibility $\tilde{\chi}(k,\omega)$ has the form:

$$\tilde{\chi}(k,\omega) = \left( -\mu\omega^2 + \alpha_S + gk^2 + wk^4 - \frac{(fk^2 - M\omega^2)^2 + 4k^2q^2\eta_S^2}{vk^4 + ck^2 - \rho\omega^2} \right)^{-1}, \quad (13)$$

The divergence of susceptibility (13) gives us the equation for phonon dispersion law $\omega(k)$:

$$\left(-\mu\omega^2 + \alpha_S + gk^2 + wk^4\right)\left(vk^4 + ck^2 - \rho\omega^2\right) - \left(fk^2 - M\omega^2\right)^2 - 4k^2q^2\eta_S^2 = 0 \quad (14)$$

Using the dimensionless wave vector $k^* = ka/\pi$, parameters (8) and introducing new dimensionless frequency $\omega^*$ and other parameters $M^*$ and $\mu^*$ in the following way:

$$\omega^* = \frac{\sqrt{4v\rho}}{c}\omega, \qquad M^* = \frac{cM}{2\rho f}, \qquad \mu^* = \frac{c\mu}{2g\rho}, \quad (15)$$

one can write the solution of biquadratic Eq.(14) in the form:

$$\omega^{*2} = -\frac{\left(k^*/a^*\right)^2\left(2\mu^* + 1 - 4F^*M^*\right) + 2\alpha_v^* + \left(k^*/a^*\right)^4\left(\mu^* + w^*\right) \pm \sqrt{Det}}{2\left(2F^*M^{*2} - \mu^*\right)}, \quad (16a)$$

$$Det = 4\left(k^*/a^*\right)^2\left(2F^*M^{*2} - \mu^*\right)\begin{pmatrix} 4\alpha_v^*\left(1 - Q^*\right) + 2\left(k^*/a^*\right)^2\left(1 - F^* + \alpha_v^*\right) \\ + w^*\left(k^*/a^*\right)^6 + \left(k^*/a^*\right)^4\left(1 + 2w^*\right) \end{pmatrix}. \quad (16b)$$

$$+ \left(\left(k^*/a^*\right)^2\left(2\mu^* + 1 - 4F^*M^*\right) + 2\alpha_v^* + \left(k^*/a^*\right)^4\left(\mu^* + w^*\right)\right)^2$$

Dispersion relation (16) contains the optic (**O**) and acoustic (**A**) phonon modes. The O mode is in fact transverse (TO), while A mode can be both longitudinal or transverse. The "gap" between these modes is proportional to the value $\frac{\sqrt{Det}}{2F^*M^{*2}-\mu^*}$, giving one the possibility to define the magnitude of the flexocoupling constants from the analytical expression.

Dependences of the dimensionless phonon frequency $\omega^*$ on the wave vector $k^*$ corresponding to the O- and A- modes are shown in **Figures 3a-b** for CF and in **Figures 3c-d** for ICF. Using the reference values from the **Table S1**, $M$ is about $\pm(1-10)\times10^{-8}$ V s$^2$/m$^2$, $\mu$ is about $10^{-18}$ s$^2$mJ and $\rho$ is about $(5-10)\times10^3$ kg/m$^3$ at normal conditions, we estimated that dimensionless parameters $M^* = (0.05 - 0.5)$ and $\mu^* = (-1.5 - +1.5)$.

Curves 1-3 in **Figures 3a** and **3c** are calculated for fixed $\alpha_v^*$ and several values of flexocoupling constant $F^*$, which signs are different for ferroics with CP and ICP due to the different sign of $g$ (see the parameters definitions (8)). Both O and A phonon dispersion curves $\omega^*(k^*)$ are virtually insensitive to $F^*$ values at $k^* \ll 1$, that reflects the gradient nature of the flexocoupling. Moreover, O modes are slightly sensitive to the values of $F^*$ for all $k^*$ values. In particular, O modes 1-3 begin to diverge very slightly with $k^*$ increasing only at $k^* \geq 0.1$. In contrast to O modes behavior, A modes are very sensitive to the $F^*$ values at $k^* \geq 0.05$. At that the behavior of A modes 1-3 calculated for different $F^*$ values is principally different for ferroics with CP and ICP. A-mode firstly bends and then disappears with $F^*$ increase at $k^* \geq 0.05$ for CF. Only for ferroics with ICP A-mode appears again at $k^* > 0.1$ and the difference between corresponding curves 1-3 decreases. Unlikely the A mode may appear at $k^* \gg 0.5$ for ferroics with CP. We cannot state this exactly, because the accuracy of the analytical expression (16) decreases with $k^*$ increase, since the higher orders of $k^*$ and its gradient should be considered in the functional (1) for $k^* \gg 0.1$.

Curves 1-3 in **Figures 3b** and **3d** are calculated for fixed flexocoupling constant $F^*$ and several values of reduced temperature $\alpha_v^*$, which signs are different for ferroics with CP (g>0) and ICP (g<0) in accordance with the parameters definitions (8). At $k^* < 0.15$ the O modes are rather sensitive to the values of $\alpha_v^*$, because soft phonons should be sensitive to the temperature changes, especially in the vicinity of the ferroic phase transition (compare the values of $\omega^*(0)$ for O modes (curves 1-3) in the **Figures 3b** and **3d**). A modes (curves 1-3) calculated for different $\alpha_v^*$ values start looking different with $k^*$ increase for both ferroics with CP and ICP. A modes calculated for ferroics with CP at different $\alpha_v^*$ begin to diverge at $k^* \geq 0.1$. Corresponding A modes for ferroics with ICP are slightly sensitive to $\alpha_v^*$ values in the region of wave vectors $0.05 \leq k^* \leq 0.15$ and insensitive outside it.

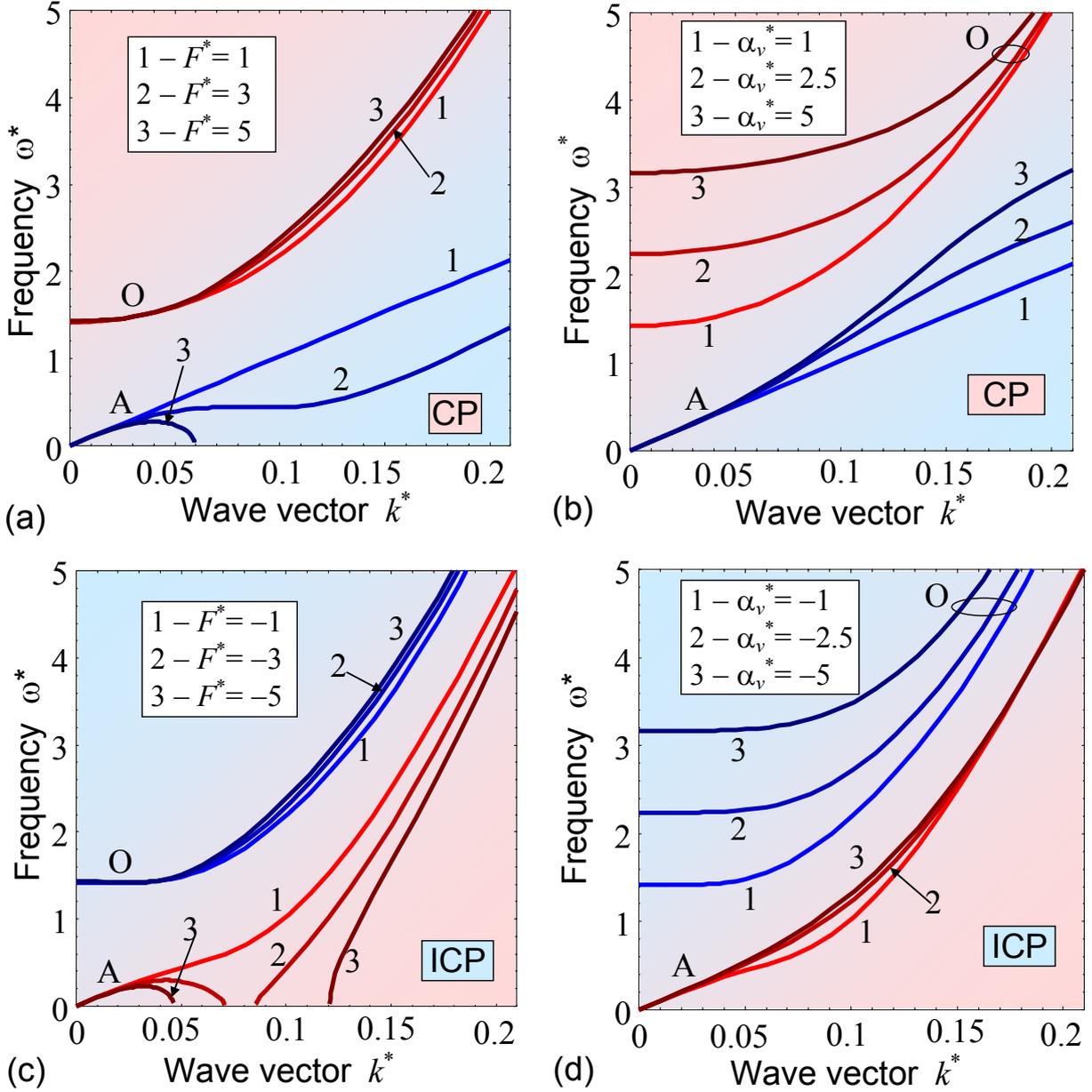

**Figure 3.** Dependences of the dimensionless phonon frequency $\omega^*$ on the wave vector $k^*$ calculated for ferroics with CP **(a,b)** and ICP **(c,d)**. Curves 1-3 in plots **(a,c)** are calculated for several values of flexocoupling constant $F^* = 1, 3, 5$ (at $\alpha_v^* = 1$) for ferroics with CP and $F^* = -1, -3, -5$ (at $\alpha_v^* = -1$) for ferroics with ICP. Curves 1-3 in plots **(b,d)** are calculated for several reduced temperature $\alpha_v^* = 1, 2.5, 5$ (at $F^* = 1$) for ferroics with CP and $\alpha_v^* = -1, -2.5, -5$ (at $F^* = -1$) for ferroics with ICP, as listed in the legends at the plots. Parameters $Q^* = 0.5$, $M^* = 0.05$, $a^* = 0.1$; $w^* = -2$, and $\mu^* = -1$ for ferroics with ICP, while $w^* = 0$ and $\mu^* = 1$ for ferroics with CP.

Obtained results demonstrate that the non-trivial differences in O and A soft phonon dispersion appeared under the change of flexocoupling constant can give the important information about influence of the flexocoupling on the SMP in ferroics with CP and ICP.

## V. Summary

Within the framework of LGD theory we studied the role of the flexocoupling between the order parameter and elastic strain gradients in the stability of SPM phases in ferroics with commensurate and incommensurate phases of the long-range order parameter. Our free energy includes the squired elastic strain gradient that is mandatory required for the system stability to arbitrary strain gradients. Performed analyses showed that the fundamental upper limit for the magnitude of the static bulk flexoelectric effect strength, established by Yudin and Tagantsev [30] under the absence of squired elastic strain gradient, should be substituted by the temperature-dependent condition on the flexoelectric coupling strength. Moreover, the temperature-dependent condition is required for the SPM phase appearance in ferroics with the higher gradients of the order parameter included.

In particular, we established that the SMP appears and becomes stable in ferroics with CP once the flexocoupling strength constant $f$ exceeds the critical value $f_{cr}$, which increases under increasing of the reduced temperature $\alpha_v$ proportional to the product of the strain gradient coefficient $v$ and temperature-dependent coefficient $\alpha \sim (T-T_c)$. For smaller $f$ the homogenous phase with the uniform distribution of the order parameter is absolutely stable. The phase diagram of ferroics with ICP plotted in coordinates $\{f, \alpha_v\}$ appeared more complex that the one for ferroics with CP only. For ferroics with ICP the SMP exists at zero $f$ until the value of $\alpha_v$ is less that the critical one $\alpha_{cr}$. When $\alpha_v > \alpha_{cr}$ the critical value $f_{cr}$ appears and increases under $\alpha_v$ increasing. Obtained analytical expressions show that $f_{cr}$ is defined by the reduced temperature, strain and order parameter gradients, striction constant and expansion coefficients on the order parameter powers in the LGD functional.

We derived analytical expressions for two modulation wave vectors $k_-$ and $k_+$ in the SMP and analyzed their dependences on parameters $f$ and $\alpha_v$ at fixed other parameters in the LGD functional. For ferroics with CP the "gap", defined as the absence of the $k_\pm$, exists for all $\alpha_v$ values and $f < f_{cr}$, while it appears only for $\alpha_v > \alpha_{cr}$ for ferroics with ICP.

We calculated the soft phonon dispersion $\omega(k)$ for ferroics with CP and ICP allowing for the squired elastic strain gradient, static and dynamic flexocoupling, and higher order gradient of the order parameter. Appeared that the dispersion $\omega(k)$ for the optic mode is slightly sensitive to the flexocoupling, while $\omega(k)$ for acoustic mode depends strongly on the coupling strength. Obtained results demonstrate that the non-trivial differences in the dispersion of optic and acoustic modes appear under the change of flexocoupling constant. Hence the phonon spectra analysis can give the important information about the influence of the flexocoupling on the SMP in ferroics with CP and ICP. These theoretical predictions require experimental verification.

## Acknowledgements

Authors are grateful to Eugene A. Eliseev for useful discussions. A.N.M. acknowledges National Academy of Sciences of Ukraine (joint Ukraine-Belarus grant 07-06-15).

## Appendix A. Tensorial form of the Helmholtz free energy and Lagrange function

LGD expansion of bulk ($F_V$) part of Helmholtz free energy $F$ on the order parameter $\eta$ and strain tensor components $u_{ij}$ have the form:

$$F = \int_V d^3r \left( \begin{array}{l} \dfrac{a_{ij}(T)}{2}\eta_i\eta_j + \dfrac{a_{ijkl}}{4}\eta_i\eta_j\eta_k\eta_l + \dfrac{a_{ijklmn}}{6}\eta_i\eta_j\eta_k\eta_l\eta_m\eta_n - \eta_i\left(E_{0i} + \dfrac{E_i^d}{2}\right) \\ + \dfrac{g_{ijkl}}{2}\left(\dfrac{\partial \eta_i}{\partial x_j}\dfrac{\partial \eta_k}{\partial x_l}\right) + \dfrac{w_{ijkl}}{2}\left(\dfrac{\partial^2 \eta_i}{\partial x_j^2}\dfrac{\partial^2 \eta_k}{\partial x_l^2}\right) + \dfrac{h_{ijk}}{2}\cdot\eta_i^2\left(\dfrac{\partial \eta_j}{\partial x_k}\right)^2 \\ -\dfrac{f_{ijkl}}{2}\left(\eta_k\dfrac{\partial u_{ij}}{\partial x_l} - u_{ij}\dfrac{\partial \eta_k}{\partial x_l}\right) - q_{ijkl}u_{ij}\eta_k\eta_l + \dfrac{c_{ijkl}}{2}u_{ij}u_{kl} + \dfrac{v_{ijklmn}}{2}\left(\dfrac{\partial u_{ij}}{\partial x_m}\dfrac{\partial u_{kl}}{\partial x_n}\right) \end{array} \right) \quad (A.1)$$

Coefficients $a_{ij}(T) = \alpha_{ij}^T(T - T_c)$ explicitly depend on temperature $T$. Coefficients $a_{ij}^S$, $a_{ijkl}$, $a_{ijkl}^S$ are supposed to be temperature independent, tensors $g_{ijkl}$, $w_{ijkl}$ and $v_{ijklmn}$ determine magnitude of the gradient energy. Tensors $v_{ijklmn}$, $w_{ijkl}$ and $a_{ijkl}$ are positively defined, $q_{ijkl}$ is the bulk striction coefficients; $c_{ijkl}$ are components of elastic stiffness tensor. The order parameter is conjugated with depolarization/demagnetization field $E_i^d$ if any exists. External field is $E_{0i}$.

Tensor $f_{ijkl}$ is the flexocoupling coefficient tensor. In fact, only the Lifshitz invariant $\dfrac{f_{ijkl}}{2}\left(\eta_k\dfrac{\partial u_{ij}}{\partial x_l} - u_{ij}\dfrac{\partial \eta_k}{\partial x_l}\right)$ is relevant for the bulk contribution. Rigorously speaking, the elastic gradient terms $v_{ijklmn}(\partial u_{ij}/\partial x_k)(\partial u_{lm}/\partial x_n)$, that is typically ignored for the ferroelectrics, are responsible for the stable smooth distribution of the order parameter at nonzero strain gradients, since the presence of Lifshitz invariant essentially changes the stability conditions and elastic boundary conditions.

Lagrange function is

$$L = \int_t dt(F - E), \quad (A.2)$$

where the kinetic energy $E$ is given by expression

$$E = \int_V d^3r \left( \dfrac{\mu}{2}\left(\dfrac{\partial \eta_i}{\partial t}\right)^2 + M_{ij}\dfrac{\partial \eta_i}{\partial t}\dfrac{\partial U_j}{\partial t} + \dfrac{\rho}{2}\left(\dfrac{\partial U_i}{\partial t}\right)^2 \right), \quad (A.3)$$

which includes the dynamic flexoelectric coupling with the tensorial strength $M_{ij}$. $U_i$ is elastic displacement and $\rho$ is the density of a ferroelectric. The strain is $u_{ij} = \dfrac{1}{2}\left(\dfrac{\partial U_i}{\partial x_j} + \dfrac{\partial U_j}{\partial x_i}\right)$.

Dynamic equations of state follows have the form of Euler-Lagrange equations allowing for the possible Khalatnikov-type relaxation of the order parameter:

$$\Gamma \frac{\partial \eta_i}{\partial t} = -\frac{\delta L}{\delta \eta_i}, \qquad \frac{\delta L}{\delta U_i} = 0. \tag{A.4}$$

## Appendix B. Static solution

In the scalar approximation free energy (A.1) acquires the following simple form for the one-component and one-dimensional case considered hereinafter:

$$F_V = \int_L dx \left( \begin{array}{c} \frac{\alpha(T)}{2}\eta^2 + \frac{\beta}{4}\eta^4 + \frac{\gamma}{4}\eta^6 - \eta E + \frac{g}{2}\left(\frac{\partial \eta}{\partial x}\right)^2 + \frac{w}{2}\left(\frac{\partial^2 \eta}{\partial x^2}\right)^2 + \frac{h}{2}\eta^2\left(\frac{\partial \eta}{\partial x}\right)^2 \\ -\frac{f}{2}\left(\eta \frac{\partial u}{\partial x} - u \frac{\partial \eta}{\partial x}\right) - qu\eta^2 + \frac{c}{2}u^2 + \frac{v}{2}\left(\frac{\partial u}{\partial x}\right)^2 \end{array} \right) \tag{B.1}$$

Thermodynamic equations of state obtained from the variation of the free energy (B.1) on order parameter $\eta$ and strain $u$ have the form:

$$\alpha\eta + \beta\eta^3 + \gamma\eta^5 - g\frac{\partial^2 \eta}{\partial x^2} + w\frac{\partial^4 \eta}{\partial x^4} - E - f\frac{\partial u}{\partial x} - 2qu\eta = 0, \tag{B.2}$$

$$cu - v\frac{\partial^2 u}{\partial x^2} + f\frac{\partial \eta}{\partial x} - q\eta^2 = 0. \tag{B.3}$$

Let us find the solution of these equations after their linearization in the vicinity of spontaneous values in k-domain $\eta = \eta_S + \int dk \exp(ikx)\tilde{\eta}$ and $u = u_S + \int dk \exp(ikx)\tilde{u}$. Perturbation field is $E = \int dk \exp(ikx)\tilde{E}$.

In a high temperature parent phase $\alpha > 0$, $\eta_S = 0$, $u_S = 0$. In a low temperature ordered phase, where $\alpha < 0$, $\eta_S \neq 0$, $u_S \neq 0$, the spontaneous strain and order parameter values can be determined from Eqs.(S.3) at zero gradients, $\alpha\eta_S + \beta\eta_S^3 + \gamma\eta_S^5 - 2qu_S\eta_S = 0$ and $cu_S - q\eta_S^2 = 0$. From here $u_S = \frac{q}{c}\eta_S^2$ and $\alpha + \beta^*\eta_S^2 + \gamma\eta_S^4 = 0$, where $\beta^* = \left(\beta - 2\frac{q^2}{c}\right)$. So that $\eta_S^2 = \frac{1}{2\gamma}\left(\sqrt{\beta^{*2} - 4\alpha\gamma} - \beta^*\right)$ for the first order phase transitions (when $\beta^* < 0, \gamma > 0$) and $\eta_S^2 = -\alpha/\beta^*$ for the second order phase transitions (when $\beta^* > 0, \gamma = 0$). Note that the condition $c\beta > 2q^2$ should be valid for the second order phase transitions.

After linearization Eqs (B.2)-(B.3) acquire the form:

$$(\alpha + 3\beta\eta_S^2 + 5\gamma\eta_S^4 - 2qu_S + gk^2 + wk^4)\tilde{\eta} - (ifk + 2q\eta_S)\tilde{u} = \tilde{E}, \tag{B.4a}$$

$$(c + vk^2)\tilde{u} + (ifk - 2q\eta_S)\tilde{\eta} = 0. \tag{B.4b}$$

The solution of these equations can be found after elementary transformations:

$$\tilde{u} = -\frac{(ifk - 2q\eta_S)}{(c + vk^2)}\tilde{\eta}, \quad \left(\alpha + 3\beta\eta_S^2 + 5\gamma\eta_S^4 - 2qu_S + gk^2 + wk^4 - \frac{4q^2\eta_S^2 + f^2k^2}{c + vk^2}\right)\tilde{\eta} = \tilde{E} \quad \text{(B.5a)}$$

In the following form:

$$\tilde{\eta} = \tilde{\chi}(k)\tilde{E}, \quad \tilde{u} = -\frac{(ifk - 2q\eta_S)}{(c + vk^2)}\tilde{\chi}(k)\tilde{E}. \quad \text{(B.6a)}$$

The generalized susceptibility (Green function) is introduced here:

$$\tilde{\chi}(k) = \left(\alpha + 3\beta\eta_S^2 + 5\gamma\eta_S^4 - 2qu_S + gk^2 + wk^4 - \frac{4q^2\eta_S^2 + f^2k^2}{c + vk^2}\right)^{-1}, \quad \text{(B.6b)}$$

The condition of the solution instability to the appearance of the spatial modulation with a period $k$ is:

$$\alpha + 3\beta\eta_S^2 + 5\gamma\eta_S^4 - 2qu_S + gk^2 + wk^4 - \frac{4q^2\eta_S^2 + f^2k^2}{c + vk^2} = 0 \quad \text{(B.7)}$$

In a low temperature ordered phase $\alpha < 0$, $\eta_S \neq 0$, $u_S \neq 0$ using that $u_S = \frac{q}{c}\eta_S^2$ and introducing the renormalized positive parameter $\alpha_S = \alpha + \left(3\beta - 2\frac{q^2}{c}\right)\eta_S^2 + 5\gamma\eta_S^4$, Eq.(B.7) becomes

$$\alpha_S + gk^2 + wk^4 - \frac{4q^2\eta_S^2 + f^2k^2}{c + vk^2} = 0. \quad \text{(B.8)}$$

Note that the parameter $\alpha_S$ is equal to $\alpha$ in a high temperature parent phase, where $\alpha > 0$, $\eta_S = 0$, $u_S = 0$, Eq.(B.8) reads $\alpha + gk^2 + wk^4 - \frac{f^2k^2}{c + vk^2} = 0$.

The equation $(\alpha_S + gk^2 + wk^4)(c + vk^2) - f^2k^2 - 4q^2\eta_S^2 = 0$ is equivalent to Eq.(B.8). After neglecting the smallest term $vwk^6$ it has the following roots:

$$k_{1,2}^\pm = \pm\sqrt{\frac{cg}{2(gv + cw)}\left(\frac{f^2}{cg} - 1 - \frac{\alpha_S v}{cg} \pm \sqrt{\left(\frac{f^2}{cg} - 1 - \frac{\alpha_S v}{cg}\right)^2 - 4\frac{(gv + cw)}{cg^2}\left(\alpha_S - 4\frac{q^2}{c}\eta_S^2\right)}\right)} \quad \text{(B.9)}$$

### Appendix C. Dynamic solution. Phonon dispersion

Dynamic equations can be obtained from the variation of the Lagrange function on the order parameter $\eta$ and displacement $U$.

$$L_V = \int_0^\infty dt \int_L dx \left(\begin{array}{l} \frac{\alpha(T)}{2}\eta^2 + \frac{\beta}{4}\eta^4 + \frac{\gamma}{6}\eta^6 - \eta E + \frac{g}{2}\left(\frac{\partial\eta}{\partial x}\right)^2 + \frac{w}{2}\left(\frac{\partial^2\eta}{\partial x^2}\right)^2 - \frac{f}{2}\left(\eta\frac{\partial^2 U}{\partial x^2} - \frac{\partial U}{\partial x}\frac{\partial\eta}{\partial x}\right) \\ -q\frac{\partial U}{\partial x}\eta^2 + \frac{c}{2}\left(\frac{\partial U}{\partial x}\right)^2 + \frac{v}{2}\left(\frac{\partial^2 U}{\partial x^2}\right)^2 - \frac{\mu}{2}\left(\frac{\partial\eta}{\partial t}\right)^2 - M\frac{\partial\eta}{\partial t}\frac{\partial U}{\partial t} - \frac{\rho}{2}\left(\frac{\partial U}{\partial t}\right)^2 \end{array}\right) \quad \text{(C.1)}$$

Euler-Lagrange equations with inclusion of Khalatnikov mechanism have the form:

$$\Gamma\frac{\partial\eta}{\partial t} + \mu\frac{\partial^2\eta}{\partial t^2} + \alpha\eta + \beta\eta^3 + \gamma\eta^5 - g\frac{\partial^2\eta}{\partial x^2} + w\frac{\partial^4\eta}{\partial x^4} - f\frac{\partial^2 U}{\partial x^2} - 2q\eta\frac{\partial U}{\partial x} + M\frac{\partial^2 U}{\partial t^2} = E, \quad \text{(C.2a)}$$

$$v\frac{\partial^4 U}{\partial x^4} + \rho\frac{\partial^2 U}{\partial t^2} - c\frac{\partial^2 U}{\partial x^2} - f\frac{\partial^2 \eta}{\partial x^2} + 2q\eta\frac{\partial \eta}{\partial x} + M\frac{\partial^2 \eta}{\partial t^2} = 0. \quad \text{(C.2b)}$$

Let us find the solution of these equations after their linearization in the vicinity of spontaneous values $\eta = \eta_S + \int dk\, \exp(ikx + i\omega t)\tilde{\eta}$ and $U = u_S x + \int dk\, \exp(ikx + i\omega t)\tilde{U}$ in k-ω domain. Perturbation field is $E = \int dk\, \exp(ikx + i\omega t)\tilde{E}$. Linearized equations (C.2) acquire the form:

$$(i\omega\Gamma - \mu\omega^2 + \alpha + 3\beta\eta_S^2 + 5\gamma\eta_S^4 - 2qu_S + gk^2 + wk^4)\tilde{\eta} + (fk^2 - 2iqk\eta_S - M\omega^2)\tilde{U} = \tilde{E}, \quad \text{(C.3a)}$$

$$(vk^4 + ck^2 - \rho\omega^2)\tilde{U} + (fk^2 + 2ikq\eta_S - M\omega^2)\tilde{\eta} = 0. \quad \text{(C.3b)}$$

The solution of these equations can be found after elementary transformations:

$$\tilde{U} = -\frac{fk^2 + 2ikq\eta_S - M\omega^2}{vk^4 + ck^2 - \rho\omega^2}\tilde{\eta}, \quad \text{(C.4a)}$$

$$\tilde{\eta} = \tilde{\chi}(k,\omega)\tilde{E} \quad \text{(C.4b)}$$

The generalized susceptibility (Green function) $\tilde{\chi}(k,\omega)$ is introduced here:

$$\tilde{\chi}(k,\omega) = \left( \begin{array}{c} i\omega\Gamma - \mu\omega^2 + \alpha + 3\beta\eta_S^2 + 5\gamma\eta_S^4 - 2qu_S + gk^2 \\ + wk^4 - \dfrac{(fk^2 - M\omega^2)^2 + 4k^2q^2\eta_S^2}{vk^4 + ck^2 - \rho\omega^2} \end{array} \right)^{-1}, \quad \text{(C.5)}$$

The condition of the solution instability to the appearance of the spatial modulation with a period $k$ is:

$$i\omega\Gamma - \mu\omega^2 + \alpha + 3\beta\eta_S^2 + 5\gamma\eta_S^4 - 2qu_S + gk^2 + wk^4 - \frac{(fk^2 - M\omega^2)^2 + 4k^2q^2\eta_S^2}{vk^4 + ck^2 - \rho\omega^2} = 0. \quad \text{(C.6)}$$

Using the solution $u_S = \dfrac{q}{c}\eta_S^2$ in Eq.(C.6a) and introducing here $\alpha_S = \alpha + \left(3\beta - 2\dfrac{q^2}{c}\right)\eta_S^2 + 5\gamma\eta_S^4$ and neglecting the damping ($\Gamma \to 0$), one can rewrite it in the form:

$$-\mu\omega^2 + \alpha_S + gk^2 + wk^4 - \frac{(fk^2 - M\omega^2)^2 + 4k^2q^2\eta_S^2}{vk^4 + ck^2 - \rho\omega^2} = 0. \quad \text{(C.7a)}$$

The equivalent form of Eq.(C.7a) is:

$$(-\mu\omega^2 + \alpha_S + gk^2 + wk^4)(vk^4 + ck^2 - \rho\omega^2) - (fk^2 - M\omega^2)^2 - 4k^2q^2\eta_S^2 = 0. \quad \text{(C.7b)}$$

In a **high temperature parent phase**, where $\alpha > 0$, $\eta_S = 0$, $u_S = 0$, the condition becomes

$$(-\mu\omega^2 + \alpha + gk^2 + wk^4)(vk^4 + ck^2 - \rho\omega^2) - (fk^2 - M\omega^2)^2 = 0 \quad \text{(C.7c)}$$

Introducing the dimensionless variables:

$$k^* = \sqrt{\frac{2v}{c}}k, \quad F^* = \frac{f^2}{cg}, \quad \alpha_v^* = \frac{\alpha_S v}{cg}, \quad w^* = \frac{cw}{vg}, \quad Q^* = 4\frac{q^2\eta_S^2}{c\alpha_S}, \quad \text{(C.8a)}$$

$$\omega^* = \frac{\sqrt{4v\rho}}{c}\omega, \quad M^* = \frac{cM}{2\rho f}, \quad \mu^* = \frac{c\mu}{2g\rho}, \quad \text{(C.8b)}$$

we rewrite Eq.(C.7a) as following:

$$\left(-\mu^*\omega^{*2} + 2\alpha_v^* + k^{*2} + \frac{w^*}{2}k^{*4}\right)(k^{*4} + 2k^{*2} - \omega^{*2}) - 2F^*(k^{*2} - M^*\omega^{*2})^2 - 4\alpha_v^* Q^* k^{*2} = 0 \quad \text{(C.9)}$$

Corresponding dispersion relation is

$$\omega^{*2} = \frac{-1}{2(2F^*M^{*2}-\mu^*)} \begin{pmatrix} k^{*2}(2\mu^*+1-4F^*M^*)+2\alpha_v^*+k^{*4}(\mu^*+w^*)\pm \\ \sqrt{\begin{matrix}4k^{*2}(2F^*M^{*2}-\mu^*)\times \\ (4\alpha_v^*(1-Q^*)+2k^{*2}(1-F^*+\alpha_v^*)+w^*k^{*6}+k^{*4}(1+2w^*)) \\ +(k^{*2}(2\mu^*+1-4F^*M^*)+2\alpha_v^*+k^{*4}(\mu^*+w^*))^2\end{matrix}} \end{pmatrix} \quad \text{(C.10)}$$

**Appendix D. Material parameters of ferroelectrics with commensurate and incommensurate phases**

Table SI. Material parameters of some ferroelectrics.

| Coefficient | PbZr$_{0.4}$Ti$_{0.6}$O$_3$ (from ref.[40] in the main text) | ICF Sn$_2$P$_2$S$_6$ (from ref. [33-35, 37-39] in the main text) | ICF Sn$_2$P$_2$Se$_6$ (from ref.[36-36] in the main text) |
|---|---|---|---|
| $\alpha_T$ ($\times 10^5$ C$^{-2}\cdot$mJ/K) | 2.12 | 16 | 16 |
| $T_C$ (K) | 691 | 337 | 193 |
| $T$ (K) | 300 | 93 | |
| $\alpha = \alpha_T(T-T_C)$ ($\times$C$^{-2}\cdot$mJ) | −828.92 (at 300 K) | −1600 (at 100 K) | positive at 300 K |
| $\beta$ ($\times 10^8$ J C$^{-4}\cdot$m$^5$) | 1.4456 | +7.42 | −4.8 |
| $\gamma$ ($\times 10^9$ J C$^{-6}\cdot$m$^9$) | 1.1154 | 35 | 85 |
| $\eta_S$ (C/m$^2$) at RT | 0.7 | 0.15 | 0 |
| $q$ ($\times 10^9$ V·m/C) | $q_{11}$=8.91, $q_{12}$=−0.787, $q_{44}$=3.18 | ~(1 – 10) exact values are unknown | exact values are unknown |
| $c$ ($\times 10^{10}$ Pa) | $c_{11}$=17.0, $c_{12}$=8.2, $c_{44}$=4.7 | ~(2 – 20) | |
| $g$ ($\times 10^{-10}$ C$^{-2}$m$^3$J) | $g_{11}$=2.0, $g_{44}$=1.0 * Estimated form the domain wall width | $g_{11}$ = −5.7, $g_{22}$ = $g_{33}$ = 5 | $g$ = − 4 |
| $h$ ($10^{-8}$ J·m$^7$/C$^4$) | 0 | 0 | 1.2 |
| $w$ ($\times 10^{-27}$ J·m$^5$/C$^2$) | 0 | 1.8 | 2.2 |
| $f$ (V) | $f_{11}$= 5, $f_{12}$= − 1, $f_{44}$= +1 *estimated from | ~ (−5 – +5) exact values are unknown | ~ (−5 – +5) exact values are unknown |
| $v$ (V s$^2$/m$^2$) | ~($10^{-7}$ – $10^{-6}$) | ~($10^{-7}$ – $10^{-6}$) | ~($10^{-7}$ – $10^{-6}$) |
| $(c/2v)^{1/2}$ (m$^{-1}$) | ~($10^9$ – $10^8$) | ~($10^9$ – $10^8$) | ~($10^9$ – $10^8$) |
| $M$ (V s$^2$/m$^2$) | $M_{11}$=6×$10^{-8}$ | unknown | unknown |
| $\mu$ ($\times 10^{-18}$ s$^2$mJ) | 1.413 | ~1 | ~1 |
| $\rho$ ($\times 10^3$ kg/m$^3$) | 8.087 * *At normal conditions | ~(5 – 10) | ~(5 – 10) |

Reference values of LDG-expansion coefficients and dimensionless parameters can be estimated for a hypothetic ferroic from the **Table S1**, namely $\alpha_S$=(0 – 2)×10$^3$×C$^{-2}\cdot$mJ, $\beta$ changes from −5×10$^8$ J C$^{-4}\cdot$m$^5$ for the ferroics with the first order phase transitions to +1×10$^8$ J C$^{-4}\cdot$m$^5$ for the ferroics with the second order phase transitions, $\gamma$ changes from 10$^9$ J C$^{-6}\cdot$m$^9$ to 10$^{11}$ J C$^{-6}\cdot$m$^9$, $\eta_S$ changes from 0 to 0.7 C/m$^2$ in dependence on temperature, $q$ is about 10$^9$ Vm/C, $c$ is about (1 – 10) ×10$^{10}$ Pa, $g$ changes from $-6\times10^{-10}$C$^{-2}$m$^3$J for ferroics with ICP to $+5\times10^{-10}$C$^{-2}$m$^3$J for ferroics with CP, $w$=0 for ferroics with CP and is about $(1 - 3)\times10^{-27}$ J·m$^5$/C$^2$ for ferroics with ICP; $f$ is about $(-5 - +5)$V, $v$ is within the range $(10^{-7} - 10^{-6})$V s$^2$/m$^2$, $M$ is about $\pm(1 - 10)\times10^{-8}$ V s$^2$/m$^2$, μ is about $10^{-18}$ s$^2$mJ and ρ is about $(5 - 10)\times10^3$ kg/m$^3$ at normal conditions. Using the reference values we estimated that dimensionless parameters changes in the range $Q^* = (0.02 - 0.8)$, $w^* = 0$ for CF $w^* = -(2 - 20)$ for ICF, $F^* = (-5 - +5)$, $v^* = (-5 - +5)$, $k^* = (0 - +0.5)$, $a^* = 0.1$, $M^* = (0.05 - 0.5)$, $\mu^* = (-1.5 - +1.5)$.